\documentclass[sn-nature]{sn-jnl}

\usepackage{graphicx}%
\usepackage{multirow}%
\usepackage{amsmath,amssymb,amsfonts}%
\usepackage{amsthm}
\usepackage{comment}%
\usepackage{mathrsfs}%
\usepackage[title]{appendix}%
\usepackage{xcolor}%
\usepackage{textcomp}%
\usepackage{manyfoot}%
\usepackage{booktabs}%
\usepackage{algorithm}%
\usepackage{algorithmicx}%
\usepackage{algpseudocode}%
\usepackage{listings}%
\usepackage{todonotes}
\usepackage{mathtools}
\usepackage{hyperref}
\usepackage{bm}
\usepackage{caption}
\usepackage[symbol]{footmisc}

\usepackage{pdfpages}

\begin{document}

\title{\vspace{-2cm}An Optimal, Universal and Agnostic Decoding Method for Message Reconstruction, Bio and Technosignature Detection}
\author[ 1,2,3,4,5]{Hector Zenil\footnote{Corresponding author. Email: \href{hector.zenil@cs.ox.ac.uk}{hector.zenil@kcl.ac.uk} }}
\author[4,5]{Alyssa Adams}
\author[3,6,7]{Felipe S. Abrah\~{a}o}
\author[3,4]{Luan Ozelim}

\affil[1]{ School of Biomedical Engineering and Imaging Sciences, King's College London, U.K.}
\affil[2]{ The Alan Turing Institute, British Library, London, U.K.}
\affil[3]{ Oxford Immune Algorithmics, Oxford, U.K.}
\affil[4]{ The Arrival Institute, U.K}
\affil[5]{ King's Institute for Artificial Intelligence, King's College London, U.K.}
\affil[6]{ Centre for Logic, Epistemology and the History of Science, University of Campinas, Brazil}
\affil[7]{ DEXL, National Laboratory for Scientific Computing, Brazil}

\abstract{
We present an agnostic signal reconstruction method for zero-knowledge one-way communication channels in which a receiver aims to interpret a message sent by an unknown source about which no prior knowledge is available and to which no return message can be sent. 
Our reconstruction method is agnostic vis-\`a-vis the arbitrarily chosen encoding-decoding scheme and other observer-dependent characteristics, such as the arbitrarily chosen computational model, probability distributions, or underlying mathematical theory. We investigate how non-random messages encode information about their intended physical properties, such as dimension and length scales of the space in which a signal or message may have been originally encoded, embedded, or generated. We focus on image data as a first illustration of the capabilities of the new method.
We argue that our results have applications to life and technosignature detection, and to coding theory in general.
}

\keywords{algorithmic information dynamics, causal deconvolution, zero-knowledge communication, technosignatures, signal processing, perturbation analysis, universal distribution, intelligent signal detection, meaning, semantics, geometry, biosignatures}



\maketitle

\section{Introduction}\label{sectionIntro}

The interpretation of any given message may be thought of as highly observer-dependent. This is because the meaning of a message can vary from one recipient to another, and indeed over time.
However, not all signals are the same. For example, signals with fewer symbols can be combined in fewer possible messages implying they are potentially restricted in the semantic space of possible meanings. Random signals with large size vocabularies may have also more ways to be combined to cover a larger semantic space.
If originally intended to have meaning by the emitter agent (or sender), truly random signals will require a lot of effort---i.e., a sufficiently high amount of computational resources to compute/decode the respective algorithmic information content expressed in its incompressible form---to produce structured messages in any semantic context at the other end of the receiver agent, otherwise they would not be (algorithmic) random. 

However, even signals such as the cosmic microwave background from all directions in the universe may later amount for the greatest scientific evidence that our universe had a beginning, as an example of possible white noise converted into a thermal signature of highly contextual meaning. 
For the past 50 years astronomers have been sweeping the skies using radio telescopes in search for signals of possible extraterrestrial intelligence origins that would be meaningful to us with no reproducible success. More recently, space missions collecting samples and data from Mars, Titan, Europa, and other extraterrestrial bodies have been focused on detecting possible biosignatures with interesting results. And some tools have been developed with different degrees of relevancy and success~\cite{abizenil,ozelim2024assemblytheoryreducedshannon,kiangPhotosynthesisAstrobiologyLooking2014, meadowsSearchHabitableEnvironments2009, nationalacademiesofsciencesengineeringandmedicineAstrobiologyStrategySearch2018,rojoCenterLifeDetection2022}. The James Webb Space Telescope, for example, has been fitted with sensors and measurement devices to scout potentially habitable solar systems \cite{fridayHowJamesWebb, haqq-misraDetectabilityChlorofluorocarbonsAtmospheres2022}. For the first time in human history, the detection of extraterrestrial biosignatures and life is at the forefront of space exploration. In this work, we ask whether some properties of messages, in special images, may be more objective even when they may have been arbitrarily constructed by their senders. In the novel \textit{Contact}, written by Carl Sagan and later made into a film, an extraterrestrial signal was received. A good deal of the story has to do with how scientists serendipitously deduce that the signal encodes an object in three dimensions, having spent months or years trying to fit it into two dimensions.




All of our knowledge about life is based on what occurs here on Earth \cite{zenillife,bedauOpenProblemsArtificial2000, clelandDefiningLife2002, dupreMetaphysicsBiology2021, kempesMultiplePathsMultiple2021, mariscalHiddenConceptsHistory2019, ruiz-mirazoUniversalDefinitionLife2004, walkerAlgorithmicOriginsLife2013, witzanyWhatLife2020}. Some researchers consider technosignature detection a more promising method of detecting life than other biosignatures because of technosignatures’ putative longevity \cite{zengauch,haqq-misraSearchingTechnosignaturesExoplanetary2022a, wrightCaseTechnosignaturesWhy2022}.
However, some researchers worry that our current theoretical toolkit for data analysis may not be sophisticated enough \cite{bottaStrategiesLifeDetection2008, enyaComparativeStudyMethods2022, neveuLadderLifeDetection2018}. 
%
%
Most of the discussion thus far has been centred around the technical justification of hardware choices and the technicalities of detection (e.g. the frequency) and not on the nature (i.e., original structural characteristics or underlying mathematical properties) of the signal itself. Thus, for example, the search for extraterrestrial signals has been mostly focused on narrow-band signals (mostly pulses) a few Hertz wide (or narrower). 
Natural cosmic noisemakers, such as pulsars, quasars, and the turbulent, thin interstellar gas of our own Milky Way do not make radio signals that are this narrow, which 
would be a reasonable justification of the bounds of the physical signal search. 
On Earth, we also find local signal-noise makers such as animals, organs, neurons, and cells. 
However, beyond identifying very basic statistical patterns, not much progress has been made on the qualitative semantic aspects of signals and communication, 
let alone on building a universal framework within which the question can be explored and analysed. 

Each different species on Earth has its own different set of sensors that it uses to interact with its environment. 
To put this in terms of signals or messages between senders and receivers, different species emit signals based on the mechanisms that allow them to do so \cite{nagelWhatItBe1974a,yongImmenseWorldHow2022}. 
They also decode messages based on their sensory capabilities, and if each species has a unique set of senses, then we can say that each species decodes the same message differently. The 52Hz whale's messages are decoded differently by humans monitoring the oceans and by other whales (who do not sense the messages at all).

One can say that if a signal is intentional, it might be decipherable. 
In order to intentionally send or receive a signal over interstellar distances, it is reasonable to assume a civilisation must understand basic science and mathematics, or at least must evolve in such a way that these are somehow encoded into a message. Hence, a message from another civilisation might use a framework similar to human-based science and mathematics to establish common ground with respect to other societies. 
Signals sent by a civilisation for its own purposes other than to also establish that ``common ground'' may be impossible for a third party to unravel, though our own mathematical tools may help.

In this paper, we introduce a method based upon the principles of (algorithmic) information theory that is aimed at sweeping over various possible encoding and decoding schemes to test our current limits on signal interpretation as we attempt to reconstruct the original partition (i.e., multidimensional space) in which the original message was given its ``meaning'' by the emitter agent (i.e., the sender intentionally aiming at communicating). Thus, we establish---for the first time, to our knowledge---a strong and deep connection between semantics, geometry and dimensional topology. 
This \emph{semantic} characteristic refers to the (algorithmic) information about the context or real-world correspondents of the emitter that the signal sent by this emitter is trying to convey to the receiver.
The real-world correspondents that we particularly investigate are images grounded in their respective contexts, that is, embedded into their respective multidimensional spaces as the emitter originally intended.
See the Sup. Inf. and \cite{Zenil2024ETpaper2} for a formal introduction to these concepts.

\emph{One-way communication} channels are those for which the receiver cannot (in principle or in practice) send any signal back to the emitter in order to help or facilitate the decoding process of the first message sent by the emitter. 
We demonstrate how signals and messages may be reconstructed by deriving the number of dimensions and the scale of each length of an object from examples of images embedded in multiple dimensions, showing a connection between irreducible information content, geometry, topology, syntax and semantics.
In addition, this article also presents results that are not only agnostic vis-\'a-vis prior knowledge of encoding-decoding schemes, but also demonstrate sufficient conditions in this \emph{zero-knowledge} scenario for enabling the reconstruction of the original \emph{message} (i.e., the original object embedded into the original multidimensional space to be conveyed to the receiver) in one-way communication channels.

\emph{Zero-knowledge communication} occurs when the receiver agent is able to correctly interpret the received signal as the originally intended message sent by the emitter agent, given that the receiver has no knowledge about the encoding-decoding scheme chosen by the emitter.
Notice that this condition of 'no prior information about the emitter agent' only applies to encoding-decoding schemes, computation models, programming languages, probability measures, underlying formal mathematical theory, and also to the original multidimensional space and object that are unknown to the receiver before any communication takes place.
However, as discussed in the Sup. Inf. and \cite{Zenil2024ETpaper2}, this zero-knowledge condition does not mean other assumptions regarding the unknown emitter agent are not being considered by the receiver agent.
For example, the assumption that the emitter agent is somehow capable of performing an arbitrary encoding and compression of the message into a signal stream; and the assumption that the sender (or emitter agent) is not trying---e.g., by employing all of its computational power to outpace that of the receiver---to deceive the receiver agent into the wrong message, as both ends of the communication channel should at least have the intention to reach the above mentioned ``common ground'' although completely independent of each other.
Also, the reader should not confuse zero-knowledge communication (ZKC) with zero-knowledge proof (ZKP), which is commonly studied in cryptography \cite{Buchanan2022Cryptographybook,Vadhan2023surveyonZKproofs,Allender2023AITstatisticalZK}.
Actually, 
one may consider ZKP and ZKC as kindred mathematical problems but as diametrically opposed counterparts with regard to the acquisition of knowledge \cite{Zenil2024ETpaper2}.

As pointed out, our method is based on the principles of Algorithmic Information Dynamics (AID)~\cite{algodyn,nmi,Abrahao2021b}, and it consists of a perturbation analysis of a received signal. 
AID is based upon the principles of information theory and the mechanisms of algorithmic probability and the universal distribution, a formal approach to a type of Artificial General/Super Intelligence that requires the massive production of a universal distribution, the mother of all models~\cite{miracle}, to build a very large (semi-computable) model of computable approximate models.  
The underlying idea is that a computable model is a causal explanation of a piece of data.  
As a result of each perturbation applied to the received signal, a new computable approximate model is built, and then compared against the observation~\cite{nmi,maininfo}.
This process then builds a large landscape of possible candidate models (along with their respective approximate complexity values) from which one can infer the best model.
In the particular context investigated in this article, the partition (i.e., multidimensional space) for which the message (image) displays sufficiently low complexity values are candidate dimensions to match the original partition that the received signal stream encodes.

Overarching frameworks such as the one introduced here can be useful not only for signal detection but also for signal deconvolution from local information in, e.g., biology (what chemical signals among cells mean).  In previous work, we showed how the same technology can be used to disentangle the 3D structure of DNA and genomic information~\cite{nar}. As introduced and formally demonstrated in~\cite{Zenil2024ETpaper2}, we show that it is an observer limitation to open-endedly approaching the theory that counter-intuitively allows the correct decoding of the sender's original encoding, but that even when subjective, the process is universal, asymptotically agnostic, mathematically optimal, and independent of computational model, programming language, or probability distributions (including either priors or updated distributions).

\section{On the information content of the Arecibo message}\label{sectionArecibo}

\begin{figure}[ht!]
	\centerline{\includegraphics[scale=0.20]{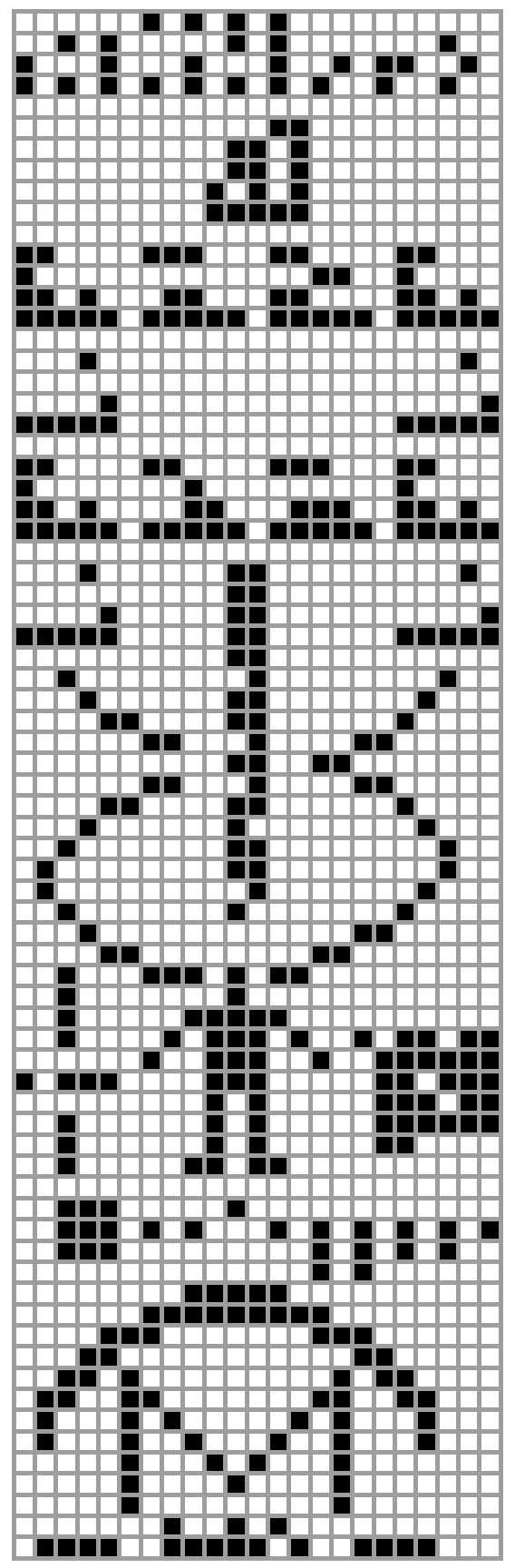}
		\hspace{1cm}
		\includegraphics[scale=0.21]{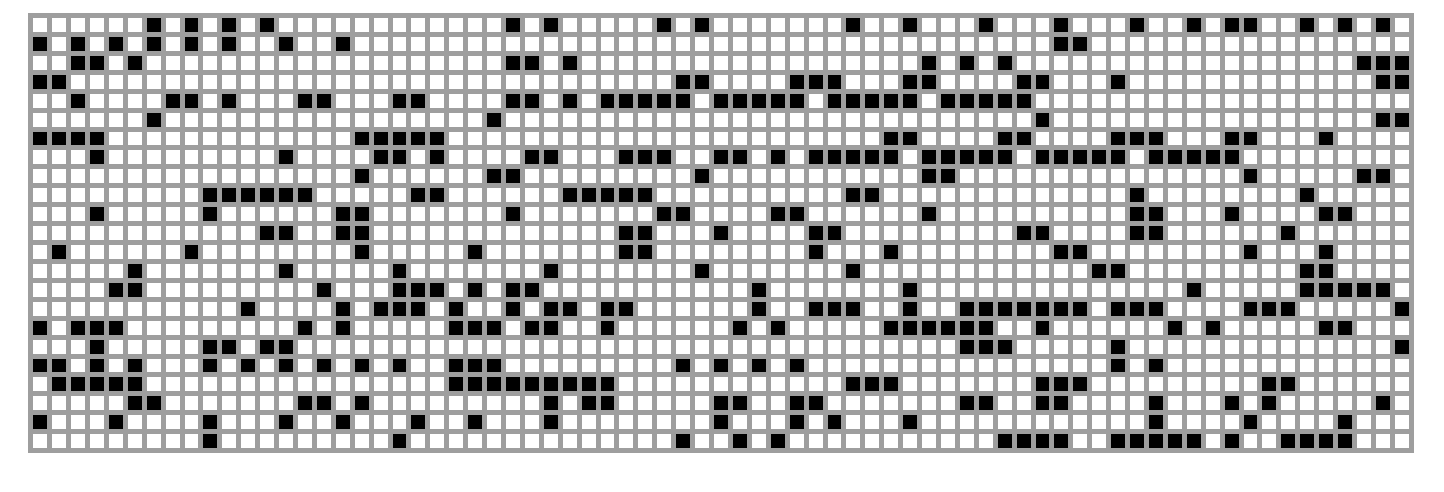}}
	\caption{\label{arecibo}\textit{Left:} The original Arecibo message intended to be reconstructed, but sent as a linear stream from the radio telescope in Arecibo, Puerto Rico. The 1,679 bits are meant to be arranged into 23 columns of 73 rows, 23 and 73 being two prime numbers which when multiplied together equal 1,679. \textit{Right:} If the stream is instead arranged into 23 rows and 73 columns, the original visual interpretations of the message are scrambled, which may result in a figure that is closer to being statistically random. What we show is that the message is still there, concealed, and can be deciphered by algorithmic deconvolution.}
\end{figure}

In 1974, the bitmap image on the left-hand side of Fig.~\ref{arecibo} was sent into space as a radio signal from the Arecibo radio telescope. In the image, there are atomic numbers for various elements and bitvectors for components of DNA. Under these, there are rough pictorial representations of a DNA molecule, a human, and the telescope. All these images seem to depend almost completely on human visual encoding/decoding conventions. On the right-hand side of Fig.~\ref{arecibo} is a reshaped version of the pattern of digits, but it is distorted so it has no obvious nested structure.  Without any sort of human context, including any indication that these are pictorial representations based on human vision, their meaning would be essentially impossible to recognise. This is especially true for message receivers who may not possess visual recognition capabilities, at least not visual capabilities that are similar to our own.

This message was reformatted from its original bidimensional state and sent as a string of 1,679 binary digits. The reasoning was that an alien civilisation with a mathematical system that receives the message will recognise 1,679 as a semi-prime number---a multiple of 23 and 73. Even if the original message could not be reconstructed in its original and intended encoding, the receiver would recognise the message as having some mathematical significance that is unlikely to be a random signal from outer space. In order to assess the information content change after structural perturbations, three metrics are measured for each perturbed array:

\begin{itemize}
    \item Entropy: Shannon entropy value of the binary array (which could be Block Entropy for higher dimensional data);
    \item BDM: The Block Decomposition Method (BDM) \cite{bdmpaper} is a strategy used to approximate the algorithmic or Kolmogorov complexity of objects like strings and graphs. Unlike conventional methods that focus on compressibility, such as lossless compression techniques, BDM divides a complex object into smaller, easier-to-handle components. These individual blocks are then evaluated based on their complexity using a reference database of small Turing machine outputs. This database provides a baseline for assessing the algorithmic content of each block. BDM builds upon the Coding Theorem Method (CTM), which relates algorithmic probability to Kolmogorov complexity by determining the chance that a random program could produce a given string on a universal Turing machine. By applying this logic to each block and summing their complexities, BDM estimates the complexity of the whole object. This approach is especially practical when calculating exact Kolmogorov complexity is too difficult due to the object’s size. By combining the complexity estimates of smaller blocks, BDM offers a broader, more refined view of complexity, surpassing the limitations of basic compressibility measures like Shannon entropy. This method reveals deeper insights into both structured and random data patterns.
    \item z-lib: the length of the b64encode of the compressed zlib version of the binary array (as a string). This is an alternative to LZW lossless compression.
\end{itemize}

The perturbations consist of rearranging the 1D stream as a 2D $m \times n$ array. Since the size of the original binary message $s=1679$ may not necessarily be reshaped as $m \times n$ (i.e., the original size is not the product of $m$ and $n$), it is important to set the maximum information loss allowed in this reshaping process. This can be set, for example as a 1 \% loss for our investigation purposes, which indicates that we are looking for $m$ and $n$ such that $mn \geq 0.99 s$, where $s$ is the length of the original encoded binary message. Also, there may be more than one pair of $m$ and $n$ values which satisfy such inequality, thus we set that $n$ is the maximum second dimension for a given $m$ value. So the dimensions actually searched are $(m,n)$, both natural numbers, such that $n=max(n_1,n_2,...n_j)$, where $m n_i \geq 0.99 s$, for $i=1,...,j$.

The following changes are then considered for each of the metrics chosen to be evaluated:

\begin{itemize}
    \item the BDM value is divided by the number of 4x4 grids used to make the partitioning of the original 2D array (i.e. $(m-3)(n-3)$);
    \item Entropy value is changed to Block Entropy, where the block size is equal to $n$;
    \item zlib b64encoding length is divided by the ratio of bits kept (i.e. $mn/s$).
\end{itemize}

\begin{figure}[ht!]
	\centering
	\includegraphics[width=0.5\textwidth]{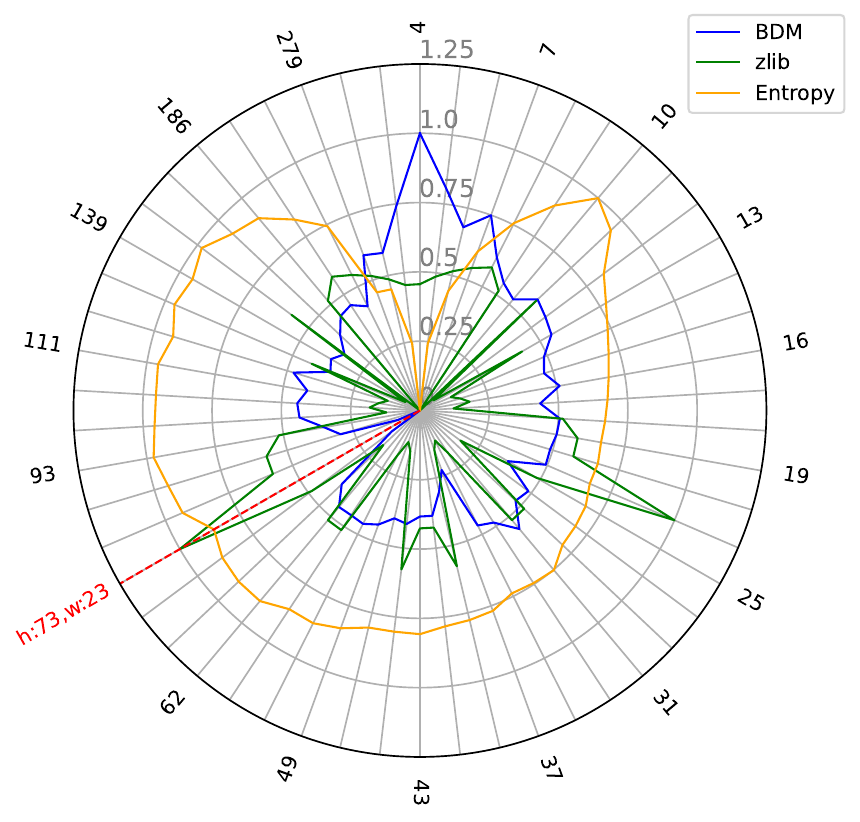}
	\caption{\label{radArecibo} Perturbations on the original 1D stream of binary digits of the Arecibo message. The ``correct'' shape of the message is represented in red and is observed for downward BDM peaks in the low-BDM region.}
\end{figure}

To make the comparisons more clear, for every metric, a MinMaxScaler was applied, such that the results range from 0 to 1. Fig.~\ref{radArecibo} presents the (structural) perturbations~\cite{Zenil2024ETpaper2} on the original Arecibo message 1D stream of binary digits. 

In general, BDM is still showing a high sensitivity to the change in information content. Block Entropy also presented a peak around the ``correct'' dimension. For small strings, z-lib may not perform well due to the overhead needed to decompress the strings in a lossless manner. Therefore, it may be of interest to upscale small strings. This could be done, for example, by changing a single bit to a square array of bits of the same value. 

\begin{figure}[ht!]
	\centerline{\includegraphics[scale=0.44]{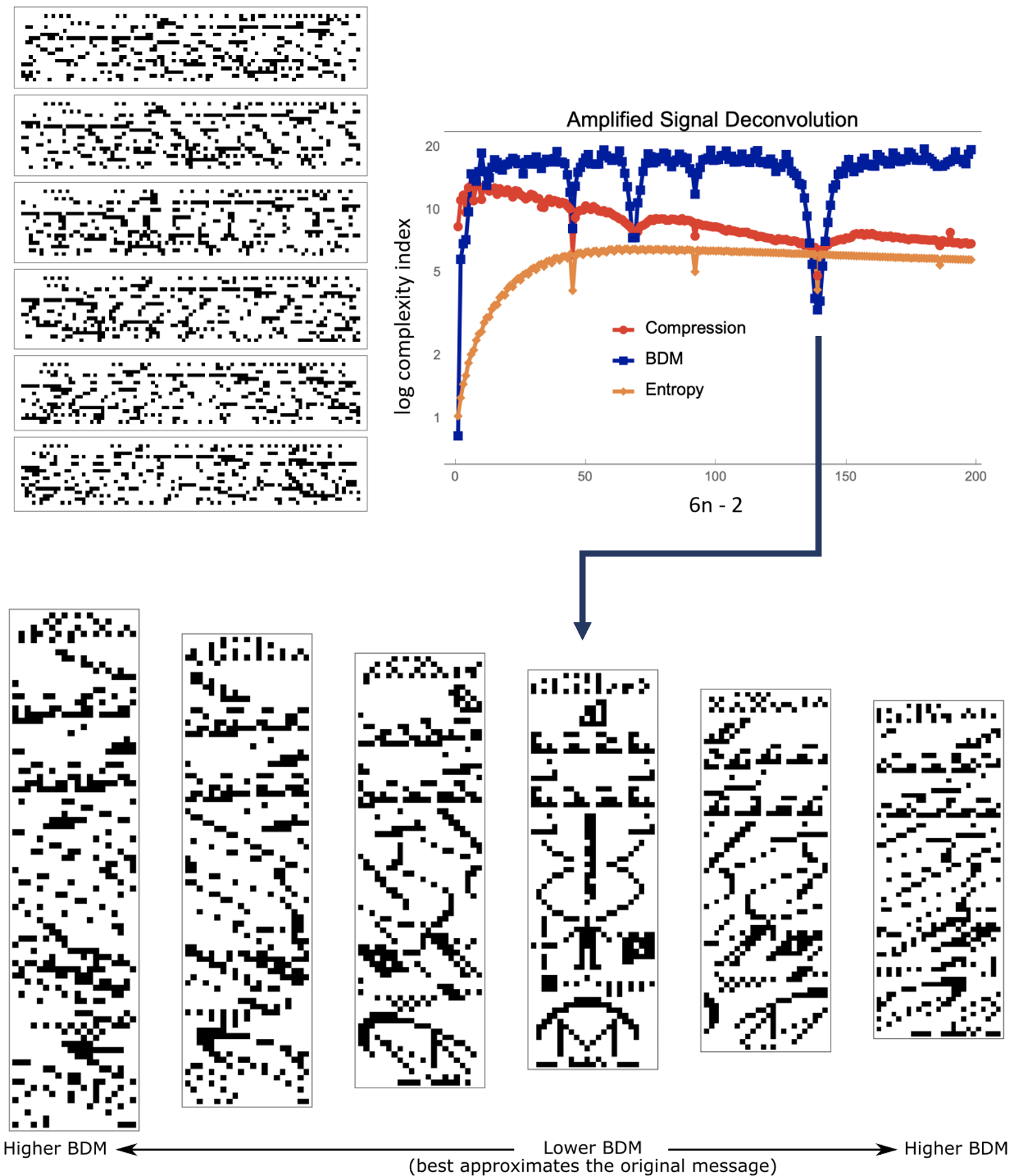}}
	\caption{\label{arecibosequence}\textit{Top left:} Most possible partitions result in random-appearing configurations with high corresponding complexity, indicating measurable randomness. \textit{Bottom:} Some partitions ($n$ values) will approximate the originally encoded meaning (third from the right). Other configurations result in images with higher complexity values. This sequence of images shows the images in the approximate vicinity of the correct bidimensional configuration (i.e., partition) and illustrates fast convergence to low complexity. \textit{Top right:} By using different information indexes across different configurations, a downward-pointing spike will indicate message (image) configurations that correspond to low-complexity image(s). In this plot, log complexity index is just a version of a scaled version of each metric, created to provide better visibility of the spikes. This allows a prior-knowledge-agnostic and objective method to infer a message's original encoding. Of the various measures, BDM, combining classical information (entropy) for long ranges and a measure motivated by algorithmic probability for short ranges, is the most sensitive and accurate in this regard. Traditional compression and entropy may also contribute to finding the right configuration amongst the top spiking candidates. The ratio of noise-to-signal was amplified in favour of the hidden structure by multiplying the original image size by 6 for both length and height (such amplification was necessary to make sure 
 compression algorithms do not bias their results from the overhead needed to decode the messages).}
\end{figure}

Fig.~\ref{arecibosequence} graphically shows how the method is robust when it comes to identifying the best way to decode a signal back to its original bidimensional space by
 sweeping over a large variety of possible dimensional configurations and measuring the resulting information content and algorithmic complexity of the candidate messages. After this process, the candidate dimension would be among the lowest algorithmic complexity configurations, under the assumption that the original message is not algorithmically random (i.e., that the original message is compressible).  
As shown in \cite{Zenil2019b} for graphs, this happens because if the original message is random, then the likelihood that a rearrangement (i.e., a new partition) will lead to a low complexity configuration is very small. On the other hand, if the original message is sufficiently compressible, 
and therefore not random, most partitions will lead to configurations of greater algorithmic randomness.
See \cite{Zenil2024ETpaper2} for the formal proofs in the general case.

\begin{figure}[ht!]
	\centering
	\includegraphics[scale=0.30]{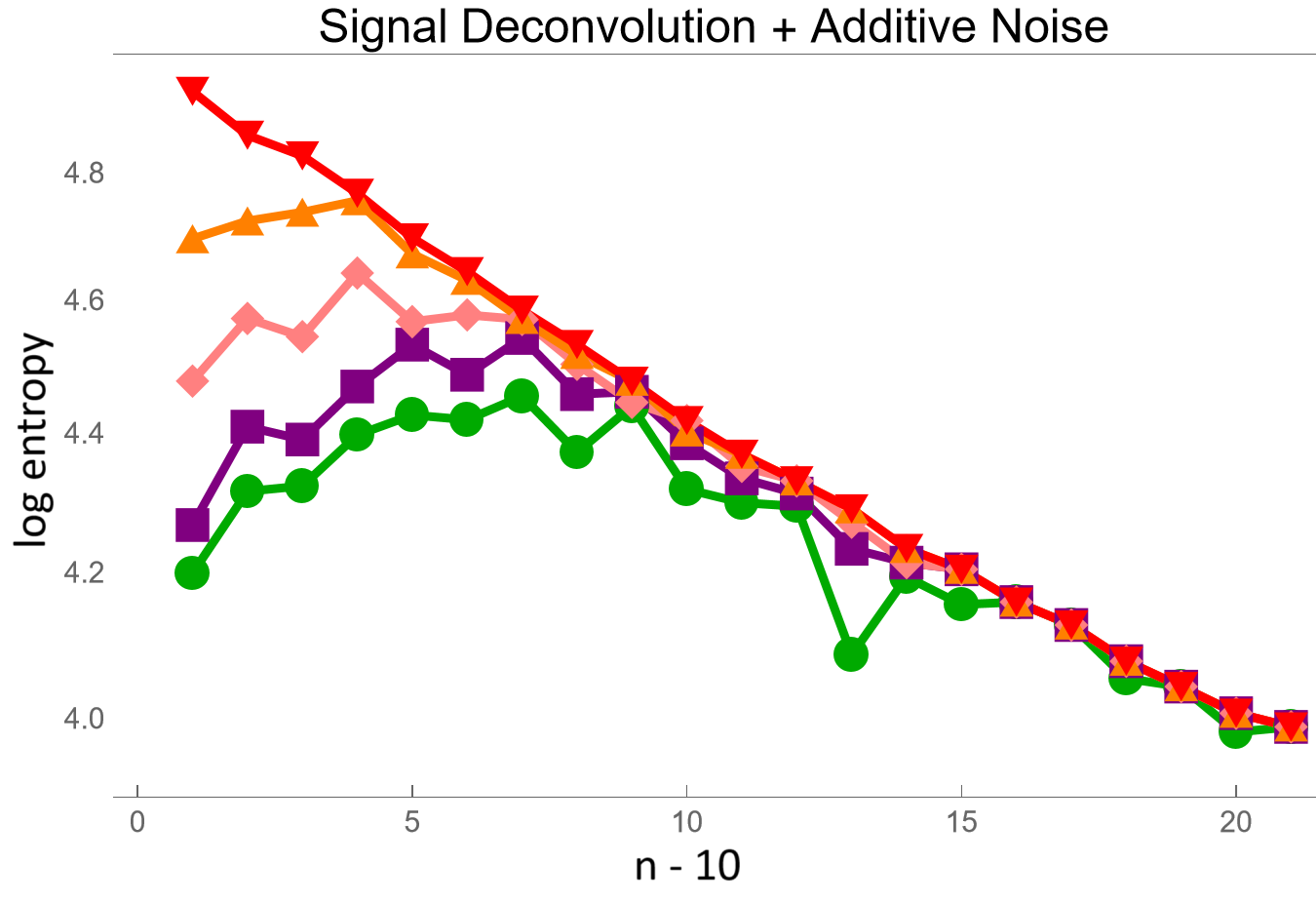}
	\caption{\label{noise}The method's resilience in the face of some noise. At 3\% of the bits of the original $23 \times 73=1679$-pixel image randomly flipped (which means about 1.5\% were binary negated), the method remains sensitive and displays a small downward spike at the 23 value, uncovering its length, but the signal gets lost when more bits are flipped.}
\end{figure}

\begin{figure}[ht!]
	\centering
	\includegraphics[scale=0.25]{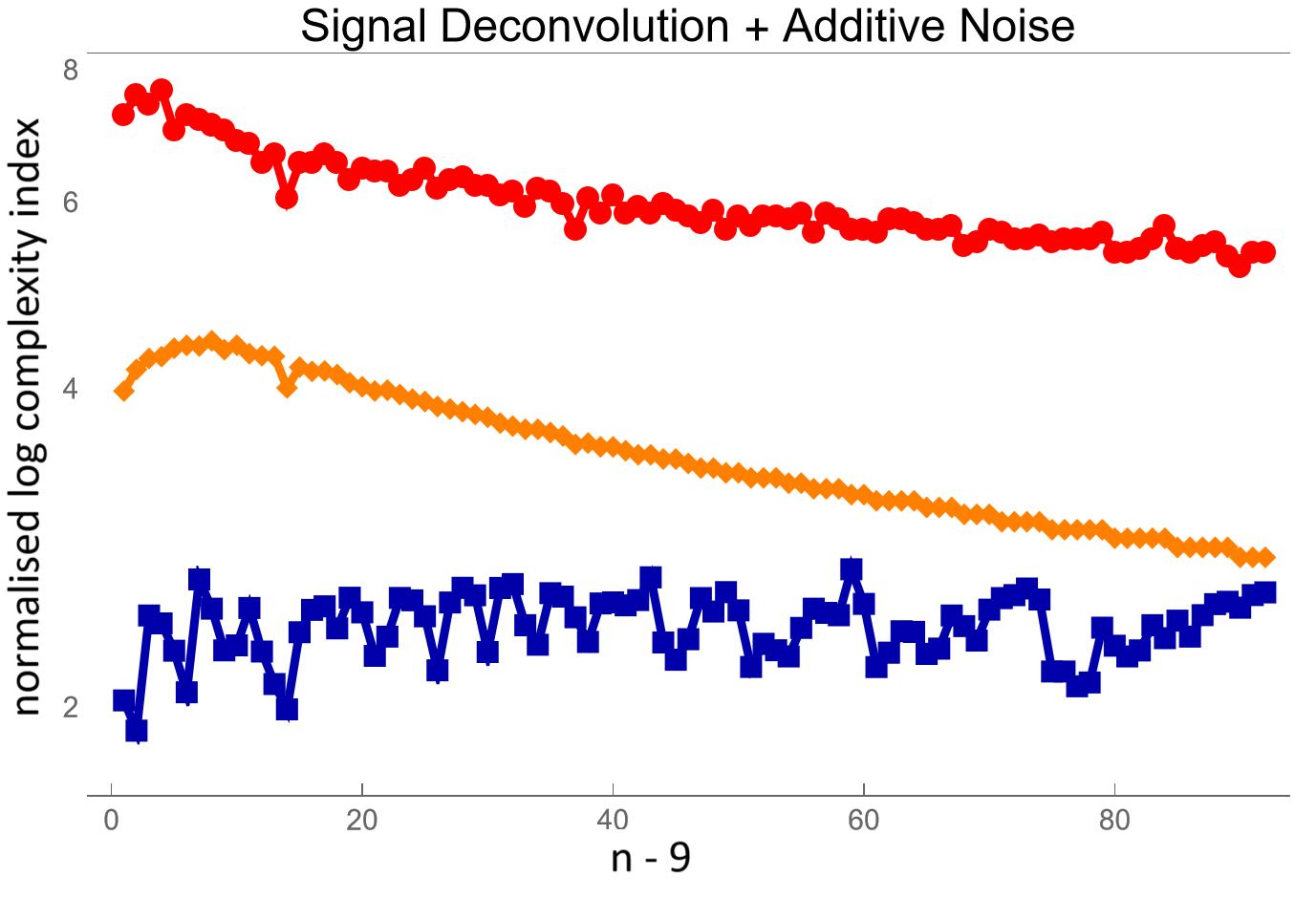}\hspace{.8cm}\includegraphics[scale=0.25]{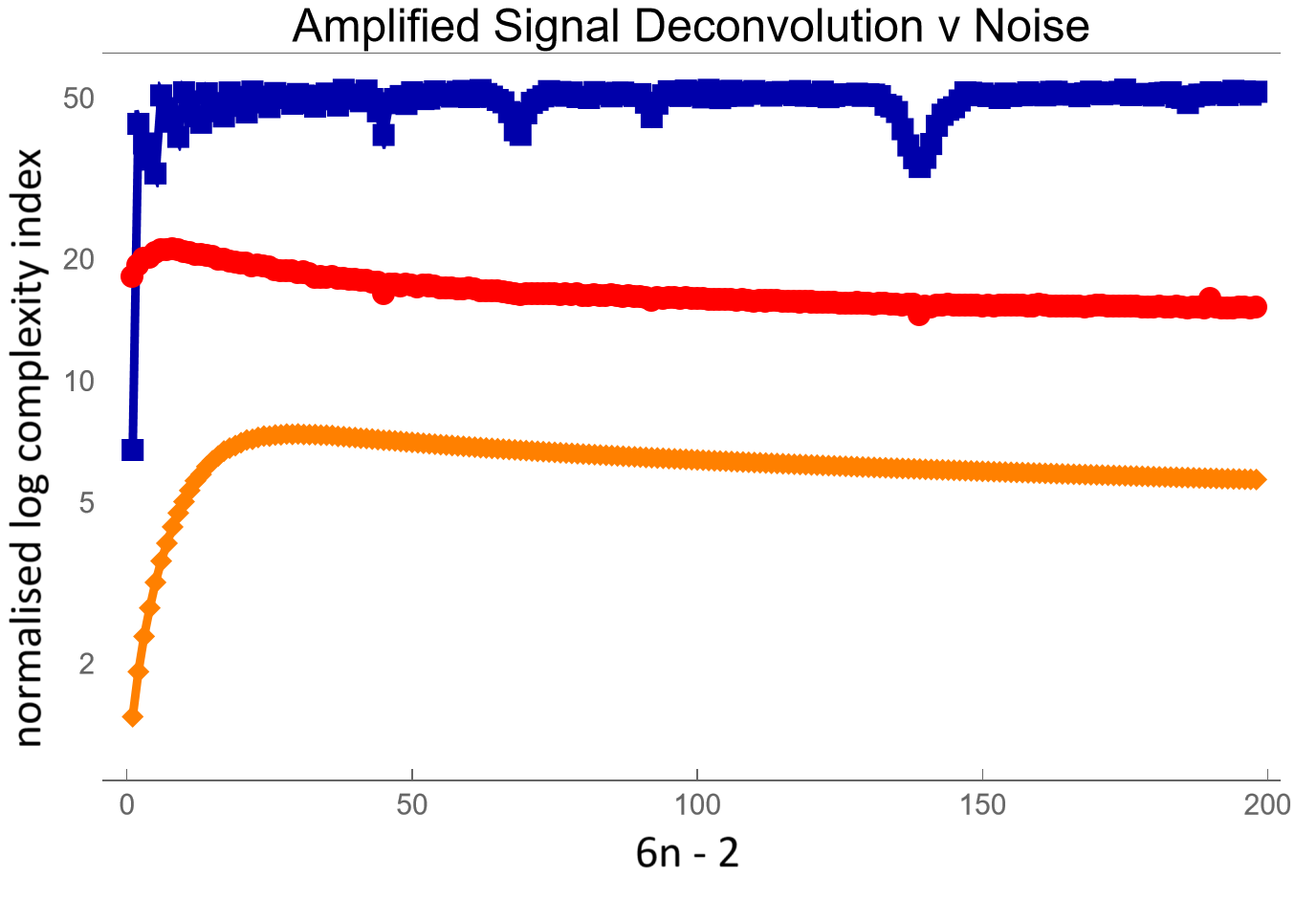}
	\caption{\label{ampnoise}After applying 3 different quantitative methods (red - z-lib Compression; blue - BDM and orange - Entropy), it is shown (left) that they are highly insensitive to signal and highly sensitive to noise (16.5\% of pixels randomly flipped). In that case, the correct partition size ($n=23$, so $n-9=14$) is modestly highlighted. However, when growing the original image (right) by a factor of 6 on each dimension (i.e. 1 pixel becomes a $6\times6$ array), the methods become less sensitive to noise and more sensitive to signal amplification, with BDM significantly outperforming Compression, and Shannon Entropy showing sensitivity at up to 60\% pixels flipped (hence about 30\% of the original image) versus Compression and Shannon Entropy, that are about 50\%
 sensitive. Downward spikes (right) are shown at $23 \times 6 - 2= 136$.}
\end{figure}

Fig.~\ref{noise} shows the method's noise resistance to indicate the candidate partition and the precise lengths of the original message. For this plot, normalised log complexity index is a scaled version of the metrics considered. Signal deconvolution before and after amplification is shown in Fig.~\ref{ampnoise}, with BDM outperforming Compression and Shannon Entropy in the face of additive noise, with Compression showing insensitivity to the original signal at about 10\% of bits flipped, and Shannon Entropy diminishing faster than BDM but slower than Compression.  In all cases, different levels of robustness at deconvolving the image dimensions are evinced.

\begin{figure}[ht!]
	\centerline{\hspace{.8cm}\includegraphics[scale=0.55]{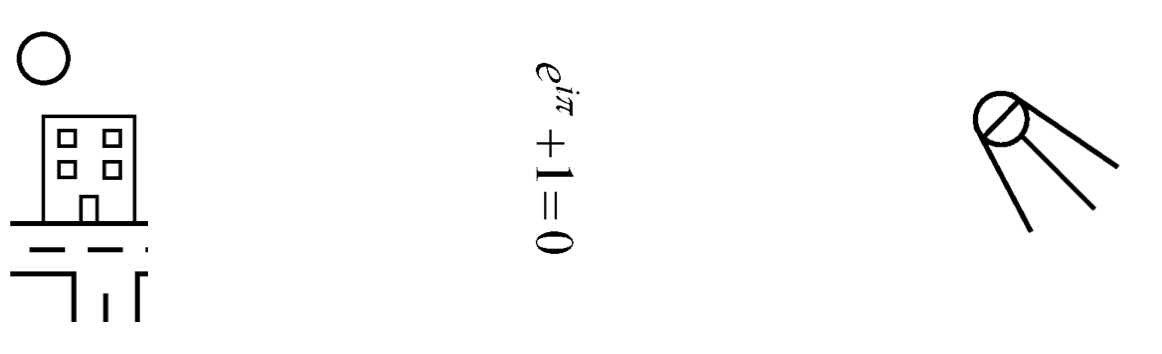}}
	
	\begin{center}
		196 \hspace{3.8cm} 82 \hspace{4cm} 215
	\end{center}
	
	\centerline{\includegraphics[scale=0.27]{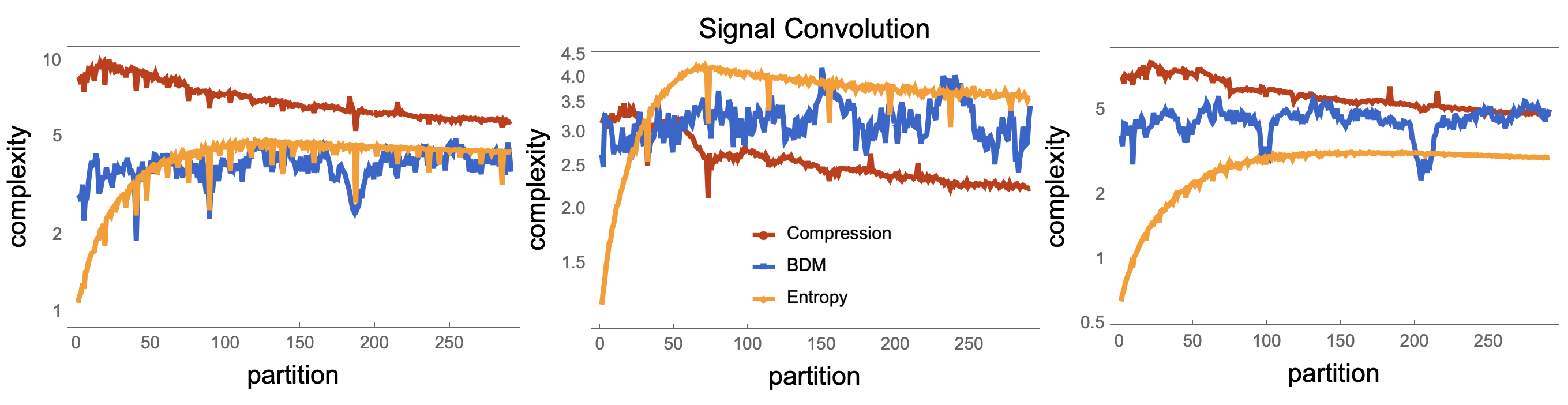}}
	
	\vspace{0.5cm}
	
	\centerline{\hspace{.8cm}\includegraphics[scale=0.37]{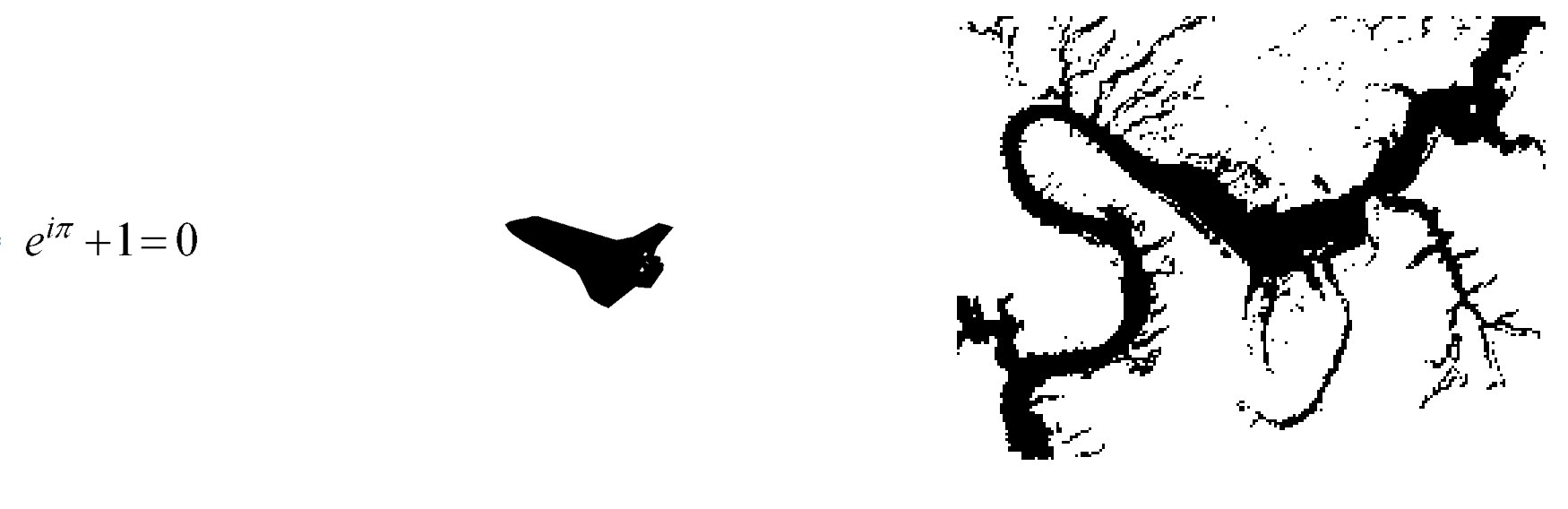}}
	
	\begin{center}
		{276 \hspace{3.6cm} 360 \hspace{3.6cm} 220}
	\end{center}
	
	\centerline{\includegraphics[scale=0.28]{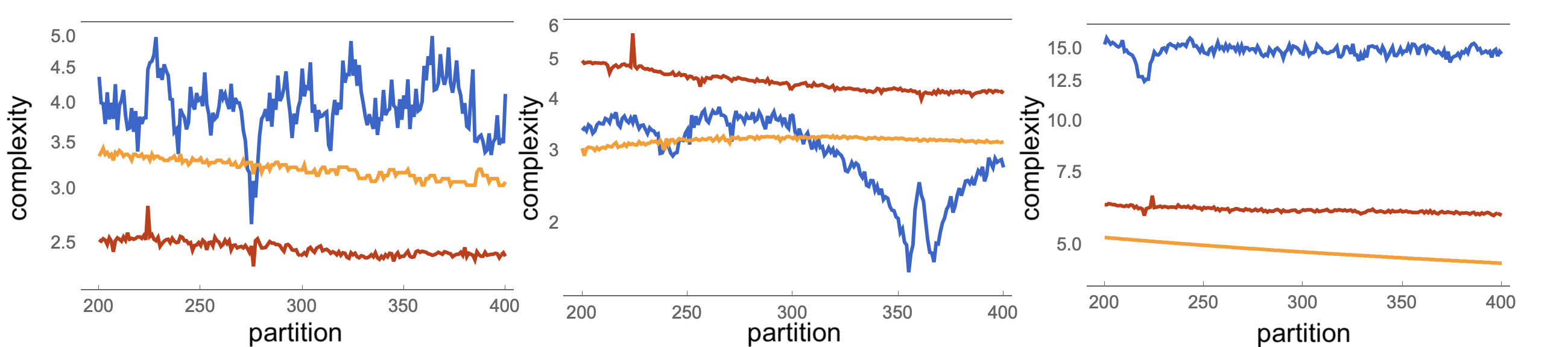}}
	\caption{\label{moreexps}Six 2D images (labelled 1 through 6 going from left to right, top to bottom) very different in nature, including a demonstration of linear-transformation invariance conforming to the underlying theory (in this case, rotation of a mathematical formula). Size invariance has actually shown amplification of signal-to-noise difference. Under each image is its correct numerical (first) dimension. The values of the three complexity indexes over possible partitions are shown below the original numerical dimension for ease of comparison. Downward spikes indicate candidates for possible original partitions. In all cases, the correct dimension value is among the top three candidates, with BDM outperforming at indicating the top candidate in 4 out of 6 cases.}
\end{figure}

Investigating this further, Fig.~\ref{moreexps} shows six bidimensional images along with their original numerical dimension. The values of the three complexity measures over possible partitions are shown below the original numerical dimension for ease of comparison. For presentation purposes, the values of BDM and z-lib b64encoded string have been divided by 1000 and then all the three metrics are shown in a logarithmic scale.

The method is invariant with regard to linear transformations such as encoding, as shown in images 2 and 4 of Fig.~\ref{moreexps}. 
Drops in complexity (downward spikes) indicate candidate dimensions for the original encoding dimensions, and thus the decoding dimension that would process the signal such that it produces the lowest-complexity message. BDM outperforms compression and entropy and correctly identifies the original message (image), encoding for 4 out of 6 images. 

\begin{figure}[ht!]
\centerline{\includegraphics[scale=0.28]{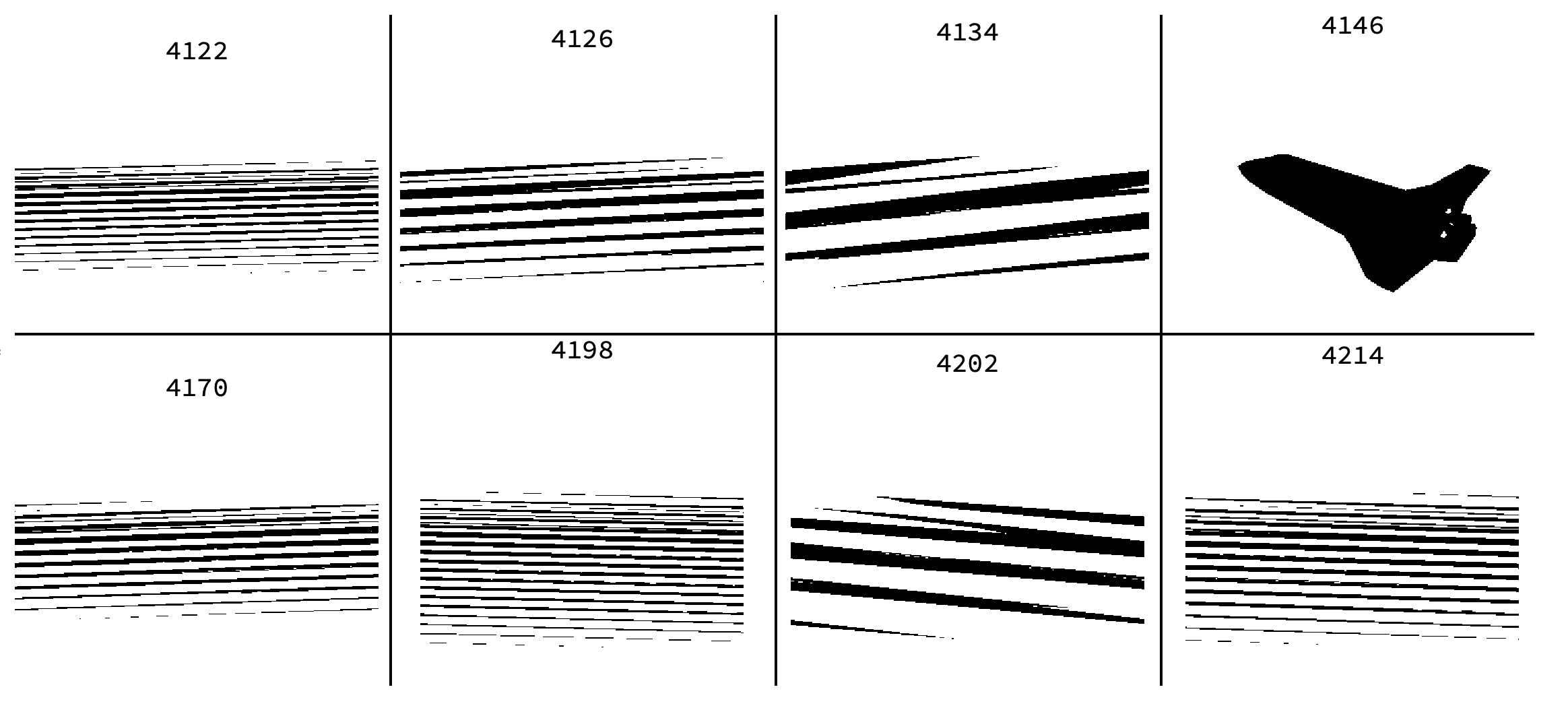}}
	\caption{\label{seqs}Sequence of lowest complexity (BDM value on top in bits) partitions of a black and white image of a space shuttle, with the top right partition as the correct reconstruction and other candidate cases in which an alignment occurs at some multiple of the correct dimension.}
\end{figure}

Some of these image decodings for the space shuttle image are shown in Fig.~\ref{seqs}. Each panel shows random realignments of the correct dimensions along with the original image (i.e., object) embedded into these new partitions, with their respective BDM values (without any scaling) displayed above each image. 
When compared to the other bidimensional spaces, the one that is most similar to the original message will have a lower complexity value.

\begin{figure}[ht!]
	
	\flushleft\footnotesize{\hspace{2.4cm}500$\times$500}
	\centerline{\includegraphics[scale=0.28]{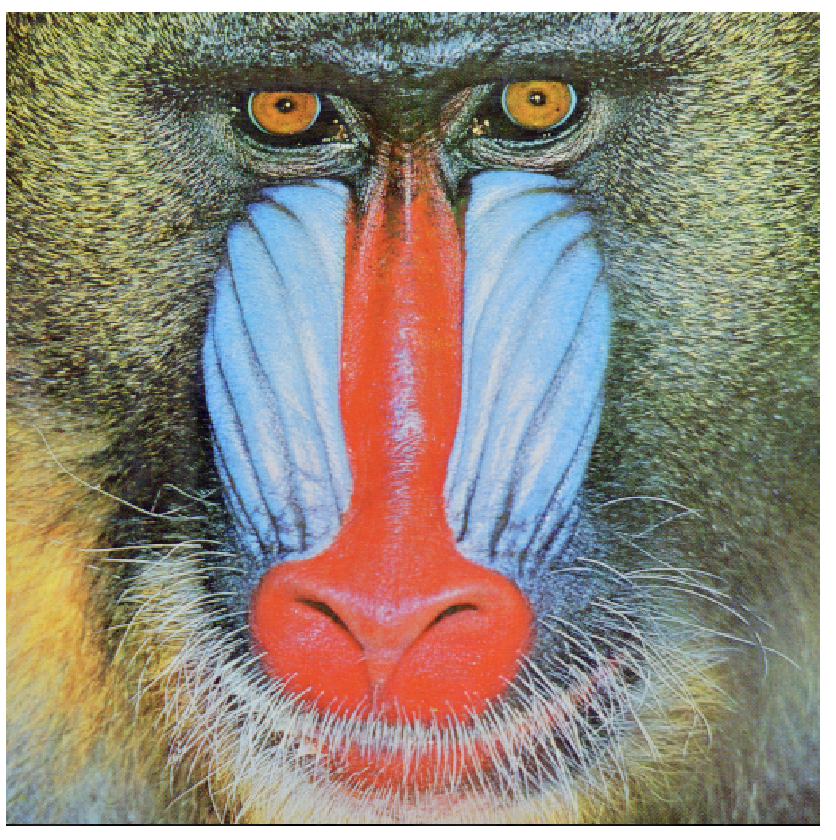}\includegraphics[scale=0.25]{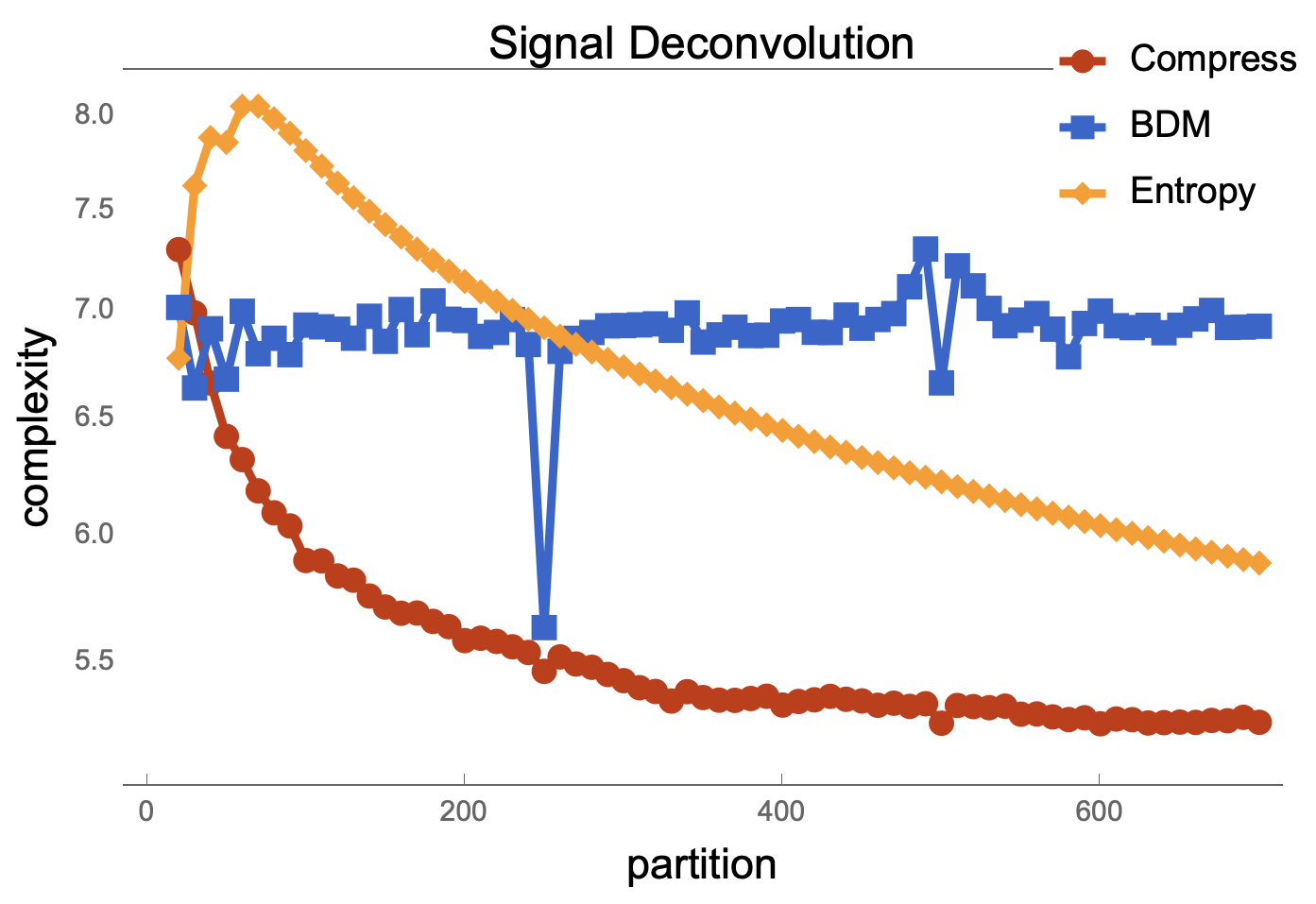}}
	\flushleft\footnotesize{\hspace{6cm}750$\times$250}
	\centerline{\includegraphics[scale=0.50]{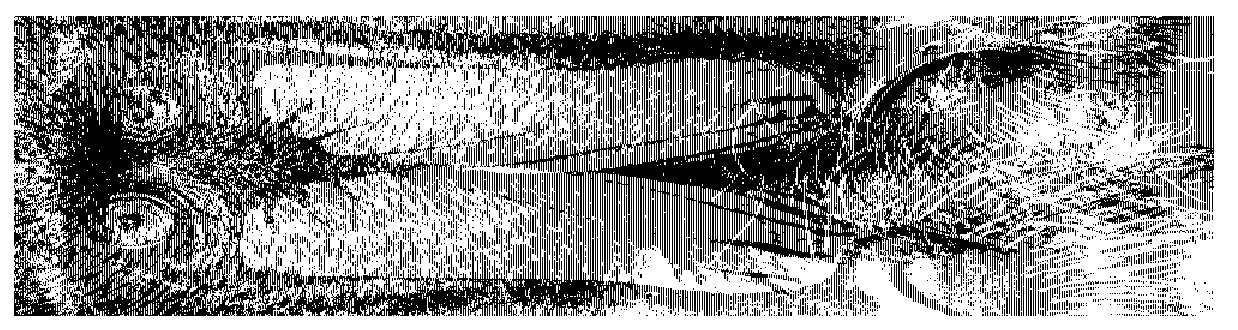}}
	\caption{\label{mandril}\textit{Top Left:} An original 2D image message and \textit{top right:} the complexity results of its various image reconstructions via linear signal decomposition (in binary, similar to how the Arecibo message was sent) into distinct partitions. BDM spikes prominently at the correct right partition with bidimensional configuration of 500 $\times$ 500 pixels, contrasting with a mirror-like image at 250 $\times$ 750 pixels (\textit{bottom}, rotated 45 degrees counterclockwise).}
\end{figure}

In translating (encoding) a colour image into binary black and white pixels, this method picks an image decoding that is similar enough to the original image (at least similar enough visually). Fig.~\ref{mandril} shows the original image on the left, the results of the method on the right, and the resulting image selected from the spike in BDM on the bottom. A mirror-like image is selected as the first candidate, followed by the correct one. This kind of spiking also suggests that the original partition can be inferred via smaller, non-random spikes at multiples or divisors of the native partition (500). In this case, small spikes at half the original partition (250) provide clues to the original encoding in 2D. 

\section{Assessing the method's performance over several different classes of images}

So far, we have illustrated the applicability of the method to a few examples of image data. In order to investigate its generalisability, we considered 102 images taken from the Caltech 101 dataset \cite{Caltech101}. Described briefly, this dataset contains pictures of objects belonging to 101 categories and an extra category named background, and the size of each image is roughly 300 x 200 pixels. 

In particular, we will test the reshaping part of the method, to check if it is capable of shortlisting the correct dimensions of each image for such a diverse set. 
As detailed in~\cite{Zenil2024ETpaper2}, we consider the upscaling of the arrays (changing a single bit to a square of size $ 6 \times  6 $---therefore a resulting $ 36 x $ total change in the image size $ x $). 
This is to make sure z-lib analysis will not be impaired by small-size samples.

Since there are several rearrangements possible, it was observed for every case studied so far that only the downward peak values are of interest. In particular, we filter only the downward peaks of BDM and its immediate left and right neighbours, selecting them if the normalised BDM value change between them and the downward peak is at most $ 1/10 $. This is done to put to the test the hypothesis that for every image, the correct dimension will be either a downward peak or its immediate neighbours.

\begin{figure}[ht!]
	\centering
	\includegraphics[scale=0.7]{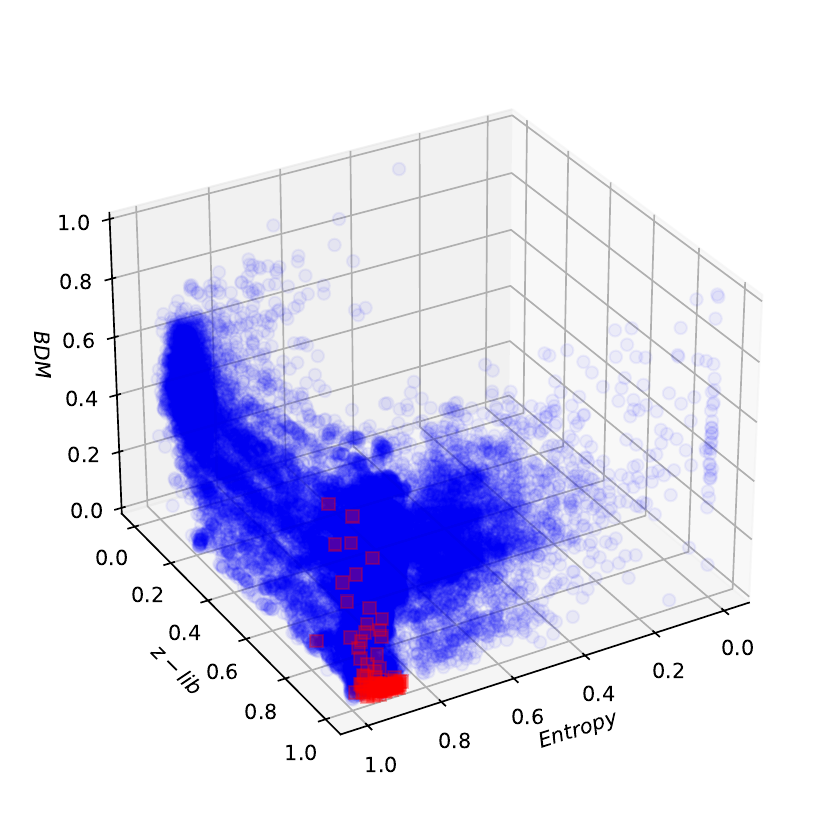}
	\caption{\label{caltech} 3D plot of the values the of three metrics for all the downward peaks and selected neighbours. Red squares, which indicate the correct dimensions of the input message, are clearly clustered in the low-normalised BDM, low-normalised compressibility (high z-lib values) and high-normalised entropy region, revealing that these three metrics, when used together, can select good candidates for such geometrical features of the input message.}
\end{figure}

After running the method on the 102 images, a total of 29511 downward BDM peaks and eligible neighbours was observed. In Fig.~\ref{caltech} the values of (BDM,Entropy,z-lib) are presented for each candidate peak, which are marked as either blue or red. The correct dimensions are the ones whose metrics are marked as red squares, while the blue circles are candidate peaks which were not correct. A very low alpha was chosen for the blue circles in order to get a clear view of the red squares.

Actually, from Fig.~\ref{caltech} it is possible to see that $ 80\% $ of the correct dimensions are the ones where the lowest BDM value is observed. Some border effects that may show up when applying BDM (which will be studied in detail in subsequent papers), could perhaps account for the fact that the other 20 \% of cases did not show up as the lowest BDM configuration. If we consider a decision tree of depth 6 to classify the peaks as either correct (1) or incorrect (0), all the peaks can be successfully separated. This decision tree is presented in the Sup. Inf..

It is important to highlight that the studies herein presented are a first approach to the problem, such that the decision tree above could not be used to extrapolate to any other image. On the other hand, it is clear that the method proposed is able to cluster candidates with correct dimensions, which empirically verifies its applicability and overall correctness.

Future studies could benefit from checking if any other metrics (or the combination of the ones already used) would be better suited to the task. As indicated, sometimes the correct dimension may not be the exact tip of the downward peak, but its immediate right or left neighbour. This may result from slight changes in the image that are not properly reflected on BDM due to the large size of the image when compared to the $ 4 \times 4 $ grid used to partition it prior to BDM calculation. Also, it could be of interest to analyse which peaks to select based on their width and prominence.

The experiments performed also revealed that when dealing with silhouettes of images, the candidate peaks are easier to classify (mostly because the background is filled with 0s). When full real images come into play, then deeper decision trees are needed.

\section{On the link between the universal distribution, AID, the method hereby proposed and Artificial General Intelligence (AGI)}

The universal distribution, also known as Levin’s distribution or the Solomonoff-Levin distribution, plays a central role in algorithmic information theory (AIT)~\cite{Chaitin2004,Calude2002,Li1997,Downey2010} by describing the optimal/maximal probability of arbitrarily generating an encodable object from a randomly generated program running on any universal Turing machine. This distribution is portrayed as a way to capture the likelihood of patterns or data occurring naturally in our universe, reflecting the intrinsic complexity (or simplicity) of objects~\cite{miracle}. The simpler a program capable of generating a particular string, the higher its probability under the universal distribution, which is maximal for any arbitrary formal-theoretic probability measure of the space of computably constructible objects, and therefore demonstrating its depth and pervasiveness in mathematics and science.

The connection between the universal distribution and AGI lies in the way AGI systems need to process and learn from diverse, complex environments. AGI is expected to generalise learning across different domains, in much the same way that the universal distribution captures the probability of generating any pattern or object. By relying on algorithmic probability, which is one of underpinnings of the universal distribution, AGI systems can prioritise more likely solutions, especially when dealing with unknown or vast search spaces.
The universal distribution aligns with AGI's requirement to balance exploration and exploitation---i.e., finding efficient solutions (simple programs with high probabilities) while remaining open to novel possibilities (complex patterns with lower probabilities). This framework supports AGI's ability to solve new, unseen problems with minimal data, mimicking human intuition, which tends to favour simpler, concise, non-redundant, and (equivalently, as demonstrated in AIT) more likely explanations based on prior experience.

In essence, the universal distribution provides a mathematical grounding for AGI's learning mechanisms by offering a way to estimate the likelihood of different hypotheses or patterns, thus guiding the system toward efficient learning and decision-making. In this scenario, the method here proposed builds on AID and on the universal distribution, using such theoretical frameworks to explore how information and complexity evolve across both space and time. AID, an extension of AIT, moves beyond algorithmic complexity by emphasising how systems dynamically adapt and transform within structured or random contexts. 
These transformations were assessed by applying perturbations to both individual elements of an input message and to its overall geometric configuration. 
Our present paper also extends AIT by not relying on the choice of a fixed mathematical space or structure~\cite{Zenil2024ETpaper2}.
Such an approach is highly relevant to AGI, as AGI systems must demonstrate the ability to flexibly adapt to novel, unforeseen challenges without prior domain-specific knowledge.

By utilising tools like the Coding Theorem Method (CTM) and the Block Decomposition Method (BDM), we can more precisely evaluate the complexity of incoming data and gain deeper insights into how an original emitter processes, stores, and generates information. AGI requires similar competencies, such as generalisation, adaptability, and the ability to learn from minimal information. Thus, both AID and the here proposed method align naturally with the core challenges of AGI research, offering a framework for intelligent systems to navigate complexity efficiently.

Moreover, the relationship between algorithmic complexity and intelligence is crucial, as AGI systems must manage complex information effectively. Techniques like BDM, which break down and assess the complexity of structured and unstructured data, provide a quantitative approach to how intelligent systems should approach learning and problem-solving. This dynamic understanding of complexity aids in the design of AGI systems capable of evolving their strategies, ensuring that they can not only handle static tasks but also adjust flexibly, mirroring human adaptability and intelligence.

\section{Conclusions}

This work advances a practical and theoretical framework that relates information, entropy, complexity, and semantics that can be extended and hence put to multiple uses in signal deconvolution, bio- and technosignature detection, cryptography, and coding theory. 
Our methods show how the receiver can decode the (multidimensional) space or structure into which the original message was sent via zero-knowledge one-way communication channels.

We believe that the present work is only one example of how our mathematical and computational framework can be used in all areas of inverse problems as an approach to a large universal generative model able to instantiate Artificial General Intelligence from first principles, particularly in application to some specific cases of message reconstruction in which prior knowledge about the source is very limited.
Such results relate information theory to fundamental areas of mathematics, such as geometry and topology, by means of compression and algorithmic probability.

\bmhead{Acknowledgements}

Felipe S. Abrah\~{a}o acknowledges support from the São Paulo Research Foundation (FAPESP), grants $2021$/$14501$-$8$ and $2023$/$05593$-$1$.

\bmhead{Author Contributions:}
 HZ conceived the theory, methods and experiments; HZ performed most of the experiments, with support from AA and LO. HZ and FA developed the theoretical framework.


\bibliography{references.bib}



\section*{Supplementary material}

# 1 Detecting technosignatures and decoding signals

Beyond Shannon's theory of information and its results in channel capacity and maximum syntactic compression, most communication research has focused on exploring the possible *syntactic* nature of signals. This includes the type of equipment, carrier, bandwidth, and medium on which that communication may happen. While alleged universal grammars have been proposed [4], they have also been criticised heavily by linguists [12]. Our current understanding of language is based on human-centric rules of communication. Questions about universal features of written, spoken, non-verbal, and internal languages remain largely unanswered beyond human communication [5, 13, 21]. Human-centred language is most certainly not similar to how neurons in a brain, cells in an organ, non-human animals, or aliens may communicate with each other.

Much of the analysis performed on message detection and biosignatures so far has been focused on chemical and molecular detection based on what we know about life here on Earth. For example, non-chemical signals from stars and other celestial bodies are compared to signals from typical stars to search for potential technosignatures [17]. Incoming radio signals as detected by SETI programs have traditionally been analysed using simple and purely statistical approaches to distinguish signals from background noise [10]. We have shown shortcomings of statistical and entropy-based approaches before [24–26] demonstrating how their limitations are enough to rethink the mathematical framework and methods for signals detection and deconvolution carrying information both of human or extraterrestrial origin.

Detecting the digits of $\pi$ in a radio signal is sometimes used as an undoubtable signal of intelligent life. However, $\pi$ is simply a property of a circle, so it is rather trivial to conceive a purely mechanical scenario that naturally generates an increasingly accurate approximation of $\pi$. For example, a geometrical probability known as Buffon's needle, is known since the 18th century. A planet or celestial body capable of emitting radio waves can encode $\pi$ in a circular orbit, if its rotational period is exactly the same as the time it takes to traverse the circle's diameter at the same speed. So, $\pi$ and other mathematical objects of low algorithmic complexity can easily be encoded in a natural mechanistic sense. However, streaming the digits of $\pi$ without specifying its ending points would appear as a stream of maximal entropy given its likely Borel normality [6] hence appearing as random looking. In fact, Ziv and Lempel showed that a sequence is normal if and only if it is incompressible by any information lossless finite-state compressor [35], meaning that any statistical compression algorithm will fail at recognising the mathematical (algorithmic) content of an object like $\pi$—content which in fact renders any finite-length initial segment of $\pi$ a highly compressible string. In order to encode $\pi$ in a radio signal, humans might map each digit in binary on a radio wave. However, the reader of such an encoded message would have to know that a property of a circle is not only first encoded into base-ten digits, but each digit is encoded in some $n$-ary encoding and its ending points.

Some researchers have suggested that the best way to get an intelligent extraterres-

trial's attention is to send it a significant numeric pattern, perhaps prime numbers or the never-ending stream of the digits of $\pi$. However, as we have discussed to the above, even these baseline assumptions can fail. Geometric-based numbers like $\pi$ can be broadly generated by natural-like (or physically plausible) processes in the universe as they can occur numerically (as an approximation) almost everywhere in physical phenomena due to their relationship to the nature of curved surfaces.

A simple Monte Carlo estimate for the value of $\pi$ can be found by generating random points on a square and counting the proportion that lies inside an inscribed circle. The probability of a point landing in the circle is proportional to the relative areas of the circle and square, so detecting $\pi$ up to several digits and attributing it to intelligent life is confounded by natural processes. A decoder might find it difficult recognising the ending points expecting $\pi$ to begin with and might just begin to encounter the segment past its starting point unless it used an algorithmic mechanism to pick up the signal in the first place.

Indeed, one way to encode $\pi$ is to send its generating mechanism, a formula, or its 2- and 3-dimensional geometric relationship among multidimensional objects. This approach can be considered closer to the one here undertaken, which partially relies on building a distribution of computable models of non-random objects like $\pi$ that can be highly compressed, and therefore identified assuming the signal received had information content like $\pi$. In fact, the present work investigates both empirically and theoretically the conditions for which a received message can be correctly decoded into the original object and its context via the estimation of the possible generating mechanisms in the first place.

For the purposes of the discussion in this article, to reconstruct the multidimensional space in which an object is embedded, the terminology "*semantic* information" is standing for the (algorithmic) information about the context or real-world correspondents of the emitter agent that the signal emitted by the very emitter agent is trying to convey. In the present work, the real-world correspondents that we particularly investigate are the objects embedded into (or grounded on) their respective multidimensional spaces (i.e., their contexts) that the emitter agent originally intended. Future research is necessary to investigate the application of our methods to other correspondent external objects, such as galaxies, stars, planets, or interstellar coordinates.

## 2 On encoding-decoding schemes, compression, and randomness

On the one hand, a non-random (i.e., compressible) message would have a large standard deviation in the range of information values for its different partitions, and thus the schema that shortens any redundancies must be read properly by an interpreter to properly translate the shorthand redundancies back into their longer form. Intuitively, this is because compression is a mapping that translates the state of the original message into a

different state aimed at minimizing any redundancies in the original messages in a shorter expression.

On the other hand, if data is already received in its most compressed form, there is little chance to recover the original multidimensional space in which the original message was embedded, let alone the original message itself. Such a message would be indistinguishable from an ideal fair-coin-toss sequence of events. This holds because of the basic properties in algorithmic information theory (AIT) and classical information theory that a message in its most compressed form is incompressible (except for a constant that does not depend on the message) by any arbitrarily chosen universal machine, programming language, or formal theory.

Indeed, it is possible for a signal stream to arrive at a naive receiver in a format that contains sufficient information about how to decompress itself. However, even in this case, the receiver must have prior knowledge about how to decode that signal into its original encoding so as to reveal the redundancies present in the original message that gave it meaning. Note that the case in which the receiver must have prior knowledge about how to decode the signal according to the emitter's previously chosen encoding scheme—so as to introduce redundancies in the original message and avoid decoding noise as meaningful information—is what already allowed the classical noisy-channel coding theorem [9] to hold.

Imagine a sender (i.e., emitter agent) encodes a flattened image as a sequence of bits and a receiver detects these bits as an incoming stream. Is there any way for the receiver to tell that the stream is a message delivered in a bidimensional array, i.e. that the message is actually an image? And, in the general case where the receiver does not know anything about the original message at all, can a lower-information transmission encode its original dimension such that the message is natively encoded? At present, there are at least two common ways to flatten an image, but we do not yet know if it is possible to reconstruct the image given any arbitrary encoding and no information about the original message or its encoding. Is this information somehow embedded in the data? If so, is it possible to recover it? As presented in [34], one can show that in the general this is not possible for objects embedded into multidimensional spaces which are defined upon a large enough number of distinct dimensions. Thus, one of the contributions of the present work is to find additional sufficient conditions for which one can avoid these impossibilities in the general case.

Without proper conditions that assure the algorithms can only pick meaningful or sound features once previously assumed that data actually contained a message, we know the presence of spurious correlations constitutes one of the dangerous of naively applying any sort of statistics-based data analysis to extract or find statistically significant patterns in sufficiently large datasets [7, 18]. A spurious correlation is a correlation found in statistical or computational analysis so that it would have occurred in a randomly generated database anyway, and therefore bears no meaningful information about the investigated phenomena nor reveals any underlying causal relation. Ideally, given a random signal,

such as white noise, no such signal interpretation should produce any meaningful message or interpretation. A random signal should not only look random, but also be interpreted as such no matter how it is encoded or decoded. Thus, along with the data compression distortions known to occur from the application of statistics-based or entropy-based methods [24, 25], the possibility of spurious findings in random messages is of major concern in decoding signal streams so as to reconstruct the original message.

These phenomena strongly suggest that for effective communication, especially for *one-way communication* between any sender and receiver, a message should not be displayed in its most compressed form, i.e. incompressible, unless the receiver has knowledge about the compression schema and can uncompress the message back into its original state (disregarding efficiency). Instead, a message should be received in an uncompressed form in first place, compressibility which is compared to the signal original size (or length, if the signal stream is unidimensional). See also [34] for the theoretical framework underpinning this intuition.

Employing perturbation analysis of irreducible information content, an area called algorithmic information dynamics (AID) [30, 31] (which is based on algorithmic information theory but applicable also to statistical approaches based on classical information) is a key asset for the search of properties of meaningful signals away from randomness due to its universal and independent (invariant in the limit) nature specially due to not being predicated upon previously given or prior probability distributions. The recognition of bit strings of various complexity values up to incompressible random sequences will require an algorithmic-information-based strategy to analyse cosmic signals. The present article shows that AID is indeed capable of harnessing the universality of algorithmic information.

Our methods in particular employ (but are independent of the chosen algorithmic complexity approximation method in the asymptotic optimal limit) an algorithm called BDM for Block Decomposition Method. BDM is a measure of algorithmic probability [26]. It in turn relies on a second algorithm called Coding Theorem Method (CTM) [20] that helps navigate model space [31]. Model space is the space of all computable models generated by exhaustively running a large number of computer programs, in this case, 5 562 153 742 336 halting programs sorted from smallest to larger on a supercomputer as a numerical approximation to the semi-computable universal model of models [20]. This fundamental model of models learns incrementally from this large computation assumed to approximate a universal distribution [15] and is trained to spike upon the closest computer program that explains the data. Other models of computation have been explored [22, 23], with some degree of correspondence [19]. BDM uses CTM to build a sequence of approximate small models that in sequence closely explain an observation or perturbation [26, 28].

Our results are not only agnostic to prior knowledge of encoding-decoding schemes, but also demonstrate sufficient conditions in this zero-knowledge context for enabling the reconstruction of the original *message* (i.e., the original object embedded into the original multidimensional space to be conveyed to the receiver) in one-way communication channels.

# 3 Basic mathematical concepts

For the purposes of the present article, we say an object has "meaning" when the object conveyed in the received signal is actually embedded into (or grounded on) the multidimensional space (i.e., the context) that the emitter agent originally intended. Each (finite and discrete) *multidimensional space* $\mathcal{S}$—no matter how complex it is—can be univocally determined by the number of dimensions, the encoding of the set of elements for each dimension, and the ordering in which each dimension appears. For example: the unidimensional space $\mathcal{S}_1$ that takes values from the natural numbers can be univocally determined by only informing the length $|\mathcal{S}_1|$ in $\mathbf{O}\left(\log\left(|\mathcal{S}_1|\right)\right)$ bits; the bidimensional space $\mathcal{S}_2$ by a pair $\left(|\mathcal{S}_{2x}|, |\mathcal{S}_{2y}|\right)$ in $\mathbf{O}\left(\log\left(|\mathcal{S}_{2x}|\right)\right) + \mathbf{O}\left(\log\left(|\mathcal{S}_{2y}|\right)\right) + \mathbf{O}(1)$ bits, where $\mathcal{S}_{2x}$ is the first dimension and $\mathcal{S}_{2y}$ is the second dimension, and so on.

In this manner, one is always able to (uniquely) decode the received message into sufficient information to completely determine the multidimensional space configuration. One example of an encoding scheme for an arbitrary multidimensional space $\mathcal{S}$ consists in encoding it in the form of a companion tuple $\boldsymbol{\tau}$ as in [2, 3]. Finite and discrete multidimensional spaces with more intricate configurations other than whole number multiplication in the form of $m_1 \times m_2 \times \cdots \times m_k$, where $m, k \in \mathbb{N}$, can, for example, be encoded by node-unaligned companion tuples $\boldsymbol{\tau}_{ua}$ as in [3, Definition 2]. Nevertheless, as demonstrated in [34], the theoretical framework of our method is agnostic vis-a-vis any arbitrarily chosen computational scheme for encoding multidimensional spaces.

Each *partition* of a $n$-bit-length linear signal stream corresponds to a distinct configuration of dimension lengths, and therefore to a distinct multidimensional space in which the total additive dimension lengths remains upper bounded by the constant $n$. For example, a picture of $4\times4$ pixels can be partitioned into other dimension-length configurations leading to different non-squared shapes, like $2\times8$, $8\times2$, or $1\times16$. Once a row (or, alternatively, a column) is fixed in a bidimensional space, the remaining rows and columns are dependent on the initial row fixation. In other words, each bidimensional partition is a single-variable dependency.

More formally, in the case of a (finite and discrete) bidimensional space $\mathcal{S}_2$, one may have different combinations of values $|\mathcal{S}_{2x}|$ and $|\mathcal{S}_{2y}|$ for each partition. However, for any partition one has it that both $1 \leq |\mathcal{S}_{2x}| \leq n$ and $\lfloor \frac{n}{|\mathcal{S}_{2x}|} \rfloor = |\mathcal{S}_{2y}|$ hold. Thus, because each partition is single-variable dependent on $\mathcal{S}_{2x}$, for any $n$-bit-length linear signal stream one can encode each of its 2D-partitions in $\mathbf{O}\left(\log(n)\right)$ bits. The same applies to 3D-partitions (or to any partition from a finite number of dimensions), except for an extra number of bits upper bounded by a partition-independent constant.

As formalised in [1] any transformation or perturbation of a finite (encoded) object into another finite (encoded) object is equivalent to an algorithmic transformation that takes the former as input and outputs the latter. That is, any transformation of a finite object into another finite object is an *algorithmic perturbation* $\mathcal{P}$ (i.e., a program) that takes the former as input and outputs the latter [1].

A (algorithmic) perturbation with low algorithmic complexity is one of these transformations where the algorithmic information content of the original object is mostly preserved under such perturbations. More formally, a *low-complexity perturbation* occurs when $\mathbf{K}(y|x) = \mathbf{O}(\log(|x|))$, where $\mathbf{K}(\cdot)$ denotes the algorithmic complexity, $x$ is the original encoded object, $y$ is the encoded object that results from the perturbation on $x$, and $|x|$ is the size of the object $x$.

Swapping *all* 0s for 1s, or swapping *all* black for white pixels in an image are examples of low-complexity perturbations (in these two cases in particular, one has it that $\mathbf{K}(y|x) = \mathbf{O}(1)$). Randomly inserting or deleting a finite number of elements (such as creating or destroying edges in a graph or flipping a finite number of bits in a string) are also examples of low-complexity perturbations [1, 31, 32]. As we saw in the above paragraphs, changing the partition is also a low-complexity perturbation of the multidimensional space into which the object $x$ is embedded. This is because any new partition can always be encoded in $\mathbf{O}(\log(|x|))$ bits. Notice that a perturbation can change either the object (e.g., by flipping bits) or its multidimensional space (e.g., by reconfiguring the dimension lengths of a given partition to achieve another distinct partition) independently, or can change both at the same time. In any of these three options, except for the identity transformation, the encoded form of the message resulting from the perturbation will be distinct from the encoded form of the original message. Both changing the partition into which the object is embedded and introducing noise into the object itself (e.g., by flipping bits) are particular examples of (algorithmic) perturbations [1, 31, 32].

We can explore the power and universality of the algorithmic coding theorem [6, 8, 11, 16] in order to investigate a lower bound for the universal (probability) distribution (given by $\mathbf{P_U}$) in comparison to an arbitrary probability measure $\mathbf{P}$. A chosen probability measure $\mathbf{P}$ may, or may not, assign a higher probability value to a correct multidimensional space. Thus, it is important to evaluate how much a universal approach such as the universal a priori probability can be wrong (i.e., diverges from $\mathbf{P}$) in the asymptotic limit. Indeed, as demonstrated in [34], the algorithmic coding theorem enables one to guarantee that the multiplicative error of

$$\mathbf{P_U}\left[\text{"event } y \text{ occurs in a given } \mathcal{S}\text{"}\right] = \mathbf{P_U}\left[y|\mathcal{S}\right]$$

with respect to

$$\mathbf{P}\left[y', \mathcal{S}'\right]$$

is bounded by the algorithmic probability of the inverse algorithmic perturbation itself. This means that the easier, or less complex, it is to retrieve $y$ embedded into $\mathcal{S}$ from $y'$ embedded into $\mathcal{S}'$ the smaller the multiplicative error of employing $\mathbf{P_U}$ instead of the arbitrarily chosen $\mathbf{P}$.

# 4 An introduction to the mathematical theory for zero-knowledge one-way communication

As shown in [31, 33], AID allows the investigation of causal analysis [28, 29] and solution of inverse problems [30] by taking advantage of a high convergence rate to the algorithmic probability brought along by BDM, probability values which remain stable across distinct models of computation, most prominently for high algorithmically probable objects.

Our theoretical results in [34] show that the method employed to achieve the empirical findings presented in the main article is sound with regard to the infinite asymptotic limit when computational resources are unbounded; and it also shows that the method can be generalised to multidimensional spaces in addition to bi- or tridimensional ones. Our inference method is general enough to remain robust vis-a-vis any arbitrarily chosen alternative method to calculate the probabilities of the events, any arbitrarily chosen encoding-decoding scheme, programming language, or computation model, without the need of assuming some bias in the possible outcomes and prior beliefs.

We demonstrate that there are conditions for which decoding the message in the original multidimensional space $\mathcal{S}$ chosen from the one in which the message has the lowest algorithmic complexity (i.e., highest algorithmic probability, as assured by the universal distribution) becomes eventually more likely to be correct than decoding the message in a distinct multidimensional space $\mathcal{S}'$, with or without noise. The underlying idea is that the larger the algorithmic complexity of the closer-to-randomness message candidate in the wrong multidimensional space with respect to the algorithmic complexity of the message in the correct multidimensional space, the larger the (algorithmic) likelihood [14, 27] of the correct multidimensional space with respect to the likelihood of the wrong closer-to-randomness multidimensional space. Thus, finding the multidimensional space that minimises the algorithmic complexity the most, eventually increases the probability of finding the correct one in comparison to the probability of finding the wrong closer-to-randomness one. Notice that this holds even if one assumes an arbitrarily different probability measure $\mathbf{P}$ as starting point.

We show that once sufficient conditions for the compressibility of objects, multidimensional spaces, and perturbations (e.g., noise) are met, one-way communication to the receiver becomes possible even if no prior knowledge about the encoding-decoding scheme chosen by the emitter is known. One can first simply look into the landscape of all possible partitions and low-complexity noise the receiver has access to while assuming no bias toward particular objects and partitions in this landscape. Then, despite having no knowledge about the encoding-decoding schemes and thus a lack of biases, one can infer with probability as high as one wishes that the message with the lowest algorithmic complexity is arbitrarily more likely to be correct than any other arbitrary computable method for assigning probabilities to message candidates.

In other words, this tells us that once certain sufficient conditions for the compressibility of objects, multidimensional spaces, and perturbations (e.g., noise) are met, one-way

communication to the receiver becomes possible even without prior knowledge about the decoding scheme chosen by the emitter. Being able to calculate the optimal values of algorithmic complexity in this case implies that one can first simply look into the landscape of all possible partitions and low-complexity noise the receiver has access to while assuming no bias toward particular objects and partitions in this landscape. Then, one can infer with probability as high as one wishes that the message with the lowest algorithmic complexity is arbitrarily more likely to be correct than any other arbitrary computable method to assign probabilities to message candidates.

In addition, although it may seem paradoxical at first glance—because one may initially think that the errors produced by the methods for approximating the algorithmic complexity would be inherited by the perturbation analysis phase itself—our theoretical and empirical results show that the perturbation analysis enables one to overcome the semi-computability limitation in algorithmic information theory, but only when one estimates with a sufficiently large amount of perturbations effected on the received signal. This is because the algorithmic complexities of the perturbed signals eventually become sufficiently larger than the complexity of the original message. Hence, a powerful enough method for approximating complexity, such as BDM, can eventually harness this divergence in order to enable the perturbation analysis to statistically distinguish the correct one (i.e., more compressible) from the wrong ones (i.e., more random). That is, an approximating method with enough computational resources can produce smaller errors than the difference of actual complexity values between the correct candidate and the perturbed ones, even though the exact complexity values remain ultimately unknown to the perturbation analyst (i.e., even though the analyst itself can never be sure it has enough computational resources in the first place).

The proofs of all theorems can be found in [34].

# 5 Separating candidate peaks and their neighbors as correct dimensions and incorrect ones

As described in the main text, a decision tree of depth 6 is able to classify the peaks as either correct (1) or incorrect (0), with a few false positives but no false negatives. The tree is by:

```
|--- BDM <= 0.00
|   |--- zlib <= 0.70
|   |   |--- class: 0
|   |--- zlib >  0.70
|   |   |--- Entropy <= 0.66
|   |   |   |--- class: 0
|   |   |--- Entropy >  0.66
|   |   |   |--- zlib <= 0.94
```

```
|   |   |   |   |   |--- Entropy <= 0.94
|   |   |   |   |   |   |--- Entropy <= 0.92
|   |   |   |   |   |   |   |--- class: 1
|   |   |   |   |   |   |--- Entropy >  0.92
|   |   |   |   |   |   |   |--- class: 1
|   |   |   |   |   |--- Entropy >  0.94
|   |   |   |   |   |   |--- class: 0
|   |   |   |--- zlib >  0.94
|   |   |   |   |--- class: 1
|--- BDM >  0.00
|   |--- BDM <= 0.25
|   |   |--- Entropy <= 0.95
|   |   |   |--- zlib <= 0.99
|   |   |   |   |--- zlib <= 0.89
|   |   |   |   |   |--- class: 0
|   |   |   |   |--- zlib >  0.89
|   |   |   |   |   |--- Entropy <= 0.90
|   |   |   |   |   |   |--- class: 0
|   |   |   |   |   |--- Entropy >  0.90
|   |   |   |   |   |   |--- class: 1
|   |   |   |--- zlib >  0.99
|   |   |   |   |--- class: 0
|   |   |--- Entropy >  0.95
|   |   |   |--- class: 0
|   |--- BDM >  0.25
|   |   |--- zlib <= 0.96
|   |   |   |--- zlib <= 0.80
|   |   |   |   |--- BDM <= 0.25
|   |   |   |   |   |--- class: 0
|   |   |   |   |--- BDM >  0.25
|   |   |   |   |   |--- class: 0
|   |   |   |--- zlib >  0.80
|   |   |   |   |--- Entropy <= 0.92
|   |   |   |   |   |--- class: 0
|   |   |   |   |--- Entropy >  0.92
|   |   |   |   |   |--- Entropy <= 0.97
|   |   |   |   |   |   |--- class: 1
|   |   |   |   |   |--- Entropy >  0.97
|   |   |   |   |   |   |--- class: 0
|   |   |--- zlib >  0.96
|   |   |   |--- class: 0
```



\end{document}